\documentclass[10pt,pre,preprint,a4paper,twocolumn,showpacs,superscriptaddress]{revtex4}
\usepackage{graphicx}

\newcommand{\be}{\begin{equation}}
\newcommand{\ee}{\end{equation}}
\newcommand{\bea}{\begin{eqnarray}}
\newcommand{\eea}{\end{eqnarray}}
\newcommand{\down}{\downarrow}
\newcommand{\up}{\uparrow}

\newcommand{\f}{\frac}

\newcommand{\bond}[1]{\tilde{#1}}

\begin{document}

\title{Directed Loop Updates for Quantum Lattice Models}

\author{Olav F. Sylju{\aa}sen}
\email{sylju@nordita.dk}
\affiliation{NORDITA, Blegdamsvej 17, DK-2100 Copenhagen {\O}, Denmark}


\date{\today}


\preprint{NORDITA-2002-71 CM}

\begin{abstract}
This article outlines how the quantum Monte Carlo directed loop update recently introduced can be applied to a wide class of quantum lattice models.
Several models are considered: spin-$s$ XXZ models with longitudinal and transverse magnetic fields, boson models with two-body interactions and 1D spinful fermion models. Expressions are given for the parameter regimes were very efficient ``no-bounce'' quantum Monte Carlo algorithms can be found.
\end{abstract}

\maketitle


\section{introduction}
The invention of non-local loop updates have made quantum Monte Carlo (QMC) simulations an indispensable tool for studying large-scale quantum many-body systems. Algorithms with non-local updates are advantageous to algorithms using only local updates. This is because they avoid the low temperature slowing down of the configuration selection process that occurs for local algorithms which severely limits the accuracy and validity of the obtained results. 

The most well-known of the non-local QMC algorithms is the Loop algorithm\cite{Evertz,evertzchapter} which can be used directly in continuous imaginary time\cite{beard} avoiding the Trotter-discretization and has proven efficient for a variety of systems. Another method is the Stochastic Series Expansion(SSE)\cite{sse1} where one relies on an expansion of the exponential in the partition function resembling more closely what is done in usual diagrammatic perturbation theory. Efficient non-local operator-loop updates in this setting was first constructed in\cite{loopsandvik}. A third method is the worm algorithm\cite{prokofev} first used to measure off-diagonal Green functions. This method is very similar to the SSE with operator-loops, but the rules described in the original formulation\cite{prokofev} for moving the worm head differs from the rules for constructing the operator-loops.     

Recently it was realized that the rules for constructing the SSE operator-loops and the rules for constructing updates in the Loop algorithm is in fact just different solutions of a set of general equations, the directed loop equations, following directly from the requirement of detailed balance. The particular setting, SSE or space-time with continuous imaginary time as used in the Loop algorithm is in fact irrelevant and a particular solution to the directed loop equations can be applied to both cases with only minor changes\cite{SS}. 
Thus the issue is not about which method is more efficient -the Loop algorithm or the SSE operator-loops. Rather, the issue is how to pick the most efficient solution to the directed loop equations. 

In Ref.~\cite{SS} the directed loop equations of the $s=1/2$ XXZ model were analyzed in detail. However as mentioned there the directed loop equations apply to a much wider class of models and we will here 
show how to construct algorithms for general lattice models by giving some general solutions to the directed loop equations. 

The directed loop equations possess often many solutions not all of them giving effective algorithms, and it will be important to choose some guidelines for how to find effective solutions. In \cite{SS} it was emphasized that the occurrences of a certain type of move, the ``bounce'' which leads to path back-tracking, effectively undoing an update already carried out, should be minimized in order for the algorithm to be effective. In \cite{SS} a region in parameter space for the $s=1/2$ XXZ model was found where bounces can be completely avoided. Here we extend this analysis to other models. However also in cases where all bounces for a given equation set can be chosen to vanish, there are for some models still choices to be made. In particular this is the case for higher spin ($s>1/2$) XXZ models\cite{HK}. Although it is impossible to test and compare the efficiency of these choices for general models we have here tested a multitude of choices for the $s=1$ Heisenberg case.

Using the solutions of the directed loop equations presented here one can construct efficient Monte Carlo moves just inputting the matrix elements of the original Hamiltonian. This is also the case for the solution of the 
directed loop equations employed in Ref.~\cite{loopsandvik}, see Ref.~\cite{henelius2}. However as we will show, the solutions used here and in Ref.~\cite{SS} lead generally to more effective algorithms.

To show the versatility of the approach we apply the rules here to several systems. Spin-s XXZ models, bosons with two-body interactions, spinful 1D fermions and the $s=1/2$ XXZ model in a transverse field. 
  
\section{The loop update}
While it was shown in Ref.~\cite{SS} that the directed loop update applies as well to the Loop algorithm as the SSE operator-loop method, we will keep the discussion within the SSE formalism here.

The starting point of the SSE method is the power series 
expansion of the partition function:
\begin{eqnarray}
Z =  {\rm Tr}\bigl\lbrace {\rm e}^{-\beta H} \bigr\rbrace
  =  \sum\limits_{\alpha}\sum\limits_{n=0}^\infty {(-\beta)^n \over n!} 
     \left \langle \alpha \left | H^n \right |\alpha \right \rangle ,
\label{zn1}
\end{eqnarray}
where the trace has been written as a sum over diagonal matrix elements in a
basis $\{|\alpha \rangle\}$. 
The Hamiltonian is written in terms of bond operators $H_b$, where
$b$ refers to a pair of sites $i(b),j(b)$,
\begin{equation}
H = - \sum_{b=1}^{N_b} H_b, \hskip6mm .
\label{hbsum}
\end{equation}
where $N_b$ is the number of bonds on the lattice. The explicit minus sign cancels the minus sign in front of $\beta$ in Eq.~(\ref{zn1}) and so if all matrix elements of $H_b$ are positive, all terms in Eq.~(\ref{zn1}) are positive. 
The bond 
operators are further decomposed into two operators;
\begin{equation}
H_b = H_{1,b} + H_{2,b},
\label{hbdef}
\end{equation}
where $H_{1,b}$ is diagonal and $H_{2,b}$ off-diagonal.

The powers of $H$ in Eq.~(\ref{zn1}) can be expressed as sums of 
products of the bond operators. Such a 
product is conveniently referred to by an operator-index sequence 
\begin{equation}
S_n = [a_1,b_1],[a_2,b_2],\ldots,[a_n,b_n],
\end{equation}
where $a_i \in \lbrace 1,2\rbrace$ corresponds to the type of operator 
($1$=diagonal, $2$=off-diagonal) and $b_i \in \lbrace 1,\ldots,N_b\rbrace$
is the bond index. Hence, 
\begin{eqnarray}
Z = \sum\limits_\alpha\sum\limits_{n=0}^\infty \sum_{S_n} 
     {\beta^n \over n!} 
     \left \langle \alpha \left | \prod_{i=1}^n H_{a_i,b_i} 
     \right | \alpha \right \rangle ,
\label{zn2}
\end{eqnarray}
where $\beta\equiv 1/T$. 
It is useful to define normalized states 
resulting when $|\alpha \rangle$ is propagated by a fraction of the SSE 
operator string: 
\begin{equation}
|\alpha (p)\rangle \sim \prod\limits_{i=1}^p H_{a_i,b_i} |\alpha\rangle .
\label{prop}
\end{equation}

\begin{figure}
\includegraphics[clip,width=8cm]{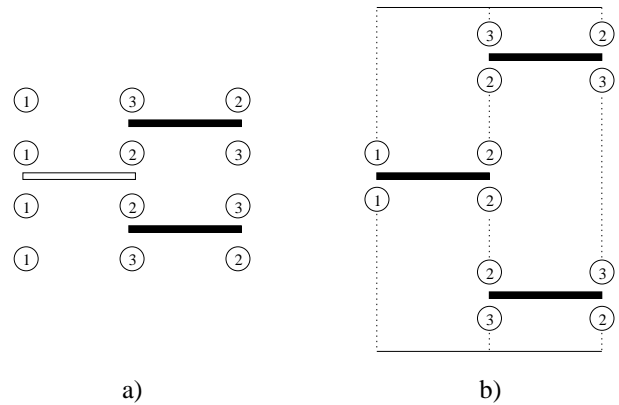}
\caption{ a) Operator-sequence for a three site system. There is one site for each column in the figure. The states on each site are labeled by encircled integers. The operators are shown as elongated boxes. There are two off-diagonal operators (filled boxes) and one diagonal operator (open box). A propagated state can be read of as one row of encircled integers. $|\alpha\rangle$ is bottom row, while the first propagated state $|\alpha(1)\rangle$ is the second row from the bottom. 
b) the operator sequence in a) shown as a vertex picture where all operators have become vertices, each with four legs. Each leg carry information about the state (encircled integer) and is connected through a dotted line to a leg on another or the same vertex. The dotted lines wrap around the top and bottom of the figure. In the vertex picture we do not distinguish between diagonal and off-diagonal operators as that is defined uniquely by the vertex legs.}
\label{vertex}
\end{figure}

In Fig.~\ref{vertex} a) a particular operator sequence is shown for a three-site system. There are three operators, one diagonal and two off-diagonal ones. The state at each site is labeled by an integer (encircled) and the propagated states can be read of as rows of encircled integers. The state $|\alpha\rangle$ is the bottom row.
For the discussion of the Monte Carlo updates it is convenient to recast the
picture in Fig.~\ref{vertex} a) into a vertex picture shown in Fig.~\ref{vertex} b) where each operator is pictured as a vertex with four legs. Each leg is carrying state information and is connected to another leg on the same or on another vertex.  

The most important part of a Monte Carlo algorithm is how the configuration is updated. In the SSE method there are two main types of updates, the diagonal update and the loop update. The diagonal update is quite trivial and inserts or removes diagonal operators in the operator string. It serves the purpose of sampling different lengths of the operator string\cite{SS}. Here we will be concerned with the loop update which changes the type of operators in the operator string, but does not change the total number of operators. In the Loop algorithm there is no notion of the diagonal update and so there the only concern is the loop update.  

The algorithm for constructing the loop update is as follows. With the
configuration mapped onto a linked vertex configuration, an initial entrance 
vertex leg is first picked at random. Then the state $s_0$ on the
entrance leg is {\em proposed} to change into a new state $s_u$ with a certain probability. An exit-leg on the starting vertex (the vertex to which the initial vertex leg belongs) is then chosen together with new states for the entrance- and exit legs according to a certain probability table. 
As will be seen below it is the solution to the directed loop equations which dictates the form of this probability table. The probability table is constructed such that the new state of the initial entrance leg is required to be equal to the proposed state $s_u$.

Changing the state on the entrance-leg or the exit-leg, or both, will result in one or two ``link-discontinuities'', where states on different legs belonging to the same link are different. A configuration with link-discontinuities does not contribute to the partition function, so the process must be repeated until the configuration has no more link-discontinuities. 

The process repeats by taking the leg connected to the exit-leg of the initial vertex as entrance leg to a new vertex, and a new exit leg and state changes are again selected according to a probability table. In order not to introduce more link-discontinuities the new state of the new entrance leg is restricted to be equal to the updated state of the previous exit leg. Thus the link-discontinuity between the previous exit leg and the current entrance leg is removed. A state change of the new exit leg will however introduce a new link-discontinuity and so the link-discontinuity is effectively moved in front of the path. 

When there is a conservation law such that the state change at the exit leg is determined by the state change at the entrance leg the link-discontinuities will only vanish when the path closes forming a loop. Then the link-discontinuity in front of the path will cancel against the discontinuity present on the link on which the initial entrance leg belongs. 
In contrast, when there is no such conservation law a link-discontinuity can vanish just because an exit state is not changed, although the entrance state {\em was} changed. One can then terminate the path if there was only one link-discontinuity present before this step. This can be achieved by requiring no link-discontinuity at the initial entrance leg\cite{henelius2}.
This starting condition is not possible when there is a conservation law as no new configuration would result, but in the absence of a conservation law, state changes on the exit leg can occur even if there is no state changes on the entrance leg.

\begin{figure}
\includegraphics[clip,width=6cm]{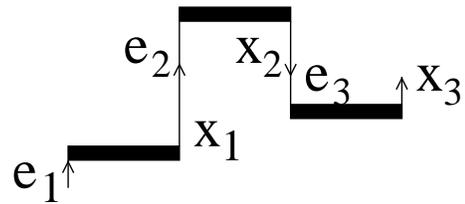}
\caption{A sequence of steps leading to a new spin configuration. The vertices are shown as horizontal lines while the legs are vertical. No states are shown, only the labeling of the different legs and the vertices they belong to.}
\label{legs}
\end{figure}

Lets now investigate how detailed balance is satisfied for the loop update. We will find the restrictions on the probabilities governing the selection of exit-legs and states as well as on the initial probabilities. This was also done in Ref.~\cite{SS}, but was restricted to the case where there is a conservation law such that the exit state is determined by the state change of the entrance leg. 
In general the detailed balance condition reads
\be
  W(s) P(s \to s^\prime) = P(s^\prime \to s) W(s^\prime) \label{detbalance}
\ee
where $W(s)$ is the weight of the configuration $s$ and $P(s \to s^\prime)$ is the
probability of changing the configuration from $s$ to $s^\prime$.
For the loop update
the probability of changing the configuration $s \equiv s^0$ to $s^\prime \equiv s^{n}$ 
can be written as a sequence of steps 
\bea
	P(s \to s^\prime) & = & \sum R(s^0,e_1) P_s(s^0(e_1)=s_0 \to s_u) 
	                        \nonumber \\ 
                    	  &   & \times P(s^0,e_1 \to s^1,x_1) 
				\nonumber \\
                     	  &   & \times P(s^1,e_2 \to s^2,x_2) 
				\nonumber \\
			  &   & \cdots	\label{updateprob} \\
	                  &   & \times P(s^{n-1},e_n \to s^{n},x_n),
	\nonumber
\eea
where $R(s^0,e_1)$ is the probability for choosing the vertex leg $e_1$ as
the initial entrance leg given the full configuration $s^0$, $P_s(s^0(e_1)=s_0 \to s_u)$ is the
probability for proposing a specific new state $s_u$ at the initial entrance leg $e_1$. The entrance(exit) leg on vertex $i$ is denoted $e_i$($x_i$). 
We denote by $s^i$ the full configuration after state changes on the $i$'th vertex in the path, so that $s^0 \equiv s$ and $s^1$ is the state obtained by possibly changing the states at $e_1$ and $x_1$. The configuration $s^n \equiv s^\prime$. The notation $s^i(j)$ refers to the {\em single  site} state at leg $j$ of the full configuration $s^i$.
$P(s^{i-1},e_i \to s^{i},x_i)$ is
the probability given the configuration $s^{i-1}$ and the entrance leg
$e_i$ on vertex $i$ to exit the same vertex at $x_i$ while changing the entrance state to $s^i(e_i)$ and the exit state to $s^i(x_i)$\cite{in1}.
In order not to introduce more link-discontinuities there are restrictions on the updated state $s^i(e_i)$ on the entrance leg: $s^i(e_i)=s^{i-1}(x_{i-1})$ for $i>1$. For $i=1$, the first vertex, $s^1(e_1)=s_u$. Thus we might as well substitute $s^1(e_1)$ for $s_u$ in Eq.~(\ref{updateprob}).
An example
illustrating some of the symbols used here is shown in Fig.~\ref{legs}. The sum in Eq.~(\ref{updateprob}) is over all paths and state changes which lead to the new configuration $s^\prime$. 

The expression for the reverse process where $s^\prime$ is changed into $s$ can easily be written down in terms of the symbols used in Eq.~(\ref{updateprob}) as {\em each} term in the sum can be reversed by just starting at the last exit leg and propagate backwards until the initial entrance leg is reached. Changing the states in opposite order brings back the original configuration $s$. The probability for the reverse process can thus be written
\bea
	P(s^\prime \to s) & = & \sum R(s^n,x_n) P_s(s^n(x_n) \to s^{n-1}(x_n)) 
	                        \nonumber \\
                    	  &   & \times P(s^{n},x_n \to s^{n-1},e_n) 
				\nonumber \\
                     	  &   & \times P(s^{n-1},x_{n-1} \to s^{n-2},e_{n-1})
				\nonumber \\
			  &   & \cdots	\label{revupdateprob} \\
	                  &   & \times P(s^1,x_1 \to s^0,e_1).
				\nonumber
\eea    
Now if one requires detailed balance to hold in the change of states on a {\em single} vertex
\bea
      & & W(s^{m-1}) P(s^{m-1},e_{m} \to s^m,x_m) \label{detbalcrit} \\
      &= & P(s^m,x_m \to s^{m-1},e_m) W(s^m), \nonumber 
\eea
it follows, by multiplying Eq.~(\ref{updateprob}) with $W(s) \equiv W(s^0)$, repeated use of Eq.~(\ref{detbalcrit}), and comparison with Eq.~(\ref{revupdateprob}) multiplied by $W(s^\prime)$ that detailed balance for the whole process is
satisfied if  
\bea
& & R(s^0,e_1) P_s(s^0(e_1) \to s^1(e_1)) \nonumber \\
&= &  \label{startcond} \\
& & R(s^n,x_n) P_s(s^n(x_n) \to s^{n-1}(x_n)) \nonumber
\eea
holds in addition to Eq.~(\ref{detbalcrit}).

When there is a conservation law, $x_n$ and $e_1$ refer to different legs on the same link and the state changes $s^0(e_1) \to s^1(e_1)$ and $s^n(x_n) \to s^{n-1}(x_n)$ occurring in Eq.~(\ref{startcond}) must therefore be opposite to each other. So if $R(s,e_1)$ is chosen to be uniform independent of $s$ and $e_1$, detailed balance requires the probabilities of opposite update proposals in the initial step to be the same.
With no conservation law we should set $P_s(s^0(e_1) \to s^1(e_1)) = \delta_{s^0(e_1),s^1(e_1)}$, causing no link-discontinuities at the initial entrance leg. 
In this case detailed balance is satisfied with a uniform $R$.
  
In addition to the above we should require that the path always exits a vertex,
\be
\sum_{x_i} \sum_{s^i(x_i)} P(s^{i-1},e_i, \to s^i,x_i) = 1, \label{sumprob}
\ee
where the sums are over all possible exit legs $x_i$ and state changes on this leg.
Note again that $s^i(e_i)$ is constrained to be equal to the exit state $s^{i-1}(x_{i-1})$ of the previous vertex. When the exit leg equals the entrance leg the entrance state is first changed, then the state on the exit leg.

Eq.~(\ref{detbalcrit}) involves only the ratio of configuration weights for configurations  which differ at most by having states changed at two legs on a single vertex. (They can also differ in the number of link-discontinuities, but these carry no weight here.) Because the full configuration weight is a product over vertex  weights it is sufficient to consider each vertex separately. 

To simplify the notation slightly we will hereafter use the notation $v$ to mean the state configuration on a {\em single} vertex. The weight of this single vertex
is denoted $W(v)$.
We also introduce $a$ as
\be
   P(v^{i-1},e_i \to v^i,x_i) = \f{a(v^{i-1},e_i \to v^i,x_i)}{W(v^{i-1})}.
\ee
Thus given the values of $a$ it is possible to construct the probability tables for how to choose exit legs and exit states. 
The equations governing the values of $a$, Eqs.~(\ref{detbalcrit}) and (\ref{sumprob}) can be written
\bea
& &      a(v^{i-1},e_i \to v^i,x_i) = a(v^i,x_i \to v^{i-1},e_i) \label{detbal1} \\
& &     \sum_{x_i} \sum_{v^i(x_i)} a(v^{i-1},e_i \to v^i,x_i) = W(v^{i-1}) \label{sumprob1}
\eea
which constitute the directed loop equations introduced in Ref.~\cite{SS} generalized to the case where there is not necessarily a conservation law dictating the state change on the exit leg.   

\section{The directed loop equations}

We will now investigate the structure of the directed loop equations, Eqs.~(\ref{detbal1}) and (\ref{sumprob1}). We will first consider situations where there is a conservation law. 

\begin{figure}
\includegraphics[clip,width=2cm]{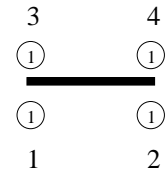}
\caption{Example of a vertex where the entrance leg is the lower left leg.}
\label{vertex1}
\end{figure}

\begin{figure}
\includegraphics[clip,width=8cm]{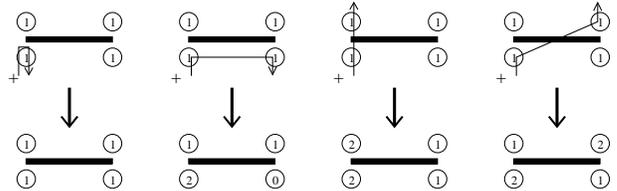}
\caption{The vertices (shown below) which results from selecting the different exit legs (shown above) in Fig.~\ref{vertex1}. The conservation law here is such that the sum of the states below each vertex equals the sum above. The $+$ on the entrance leg indicates that the state on the inleg is to be increased by unity}
\label{vertex2}
\end{figure}

In order to describe the general form of the directed loop equations
for a general interaction with $N_{\rm legs}$ legs it is convenient to abbreviate the labeling 
somewhat from that used in the previous section. To define this new labeling,
we start by selecting a reference vertex (which can be any of the allowed
vertices) and label its weight $W_1$. We then choose an entrance leg and 
label this leg as leg $1$, and then number the rest of the legs on this 
vertex $2,3$ ... $n=N_{\rm legs}$, see Fig.~\ref{vertex1} for an example of a two-site interaction. 
We then pick a specific way of changing
the state at the entrance leg. Thus the equations derived will apply
to the vertex with weight $W_1$ and with the specific state change at entrance leg $1$.  
On changing the states at both the entrance and exit legs (according to the conservation law) one arrives at a 
new vertex. A specific example is shown in Fig.~\ref{vertex2}.
Distributing the weight over all possible exit legs 
according to Eq.~(\ref{sumprob1}) gives
\be
        W_1 = a_{11} + a_{12} + \cdots + a_{1n},
\ee
where we have labeled the weights $a_{ij}$ by their
entrance ($i$) and exit ($j$) legs. 
Now label the weight of the 
vertex reached by exiting at leg $j$ as $W_j$. Thus if the exit was on leg 
$2$ we would label that vertex $W_2$. 

\begin{figure}
\includegraphics[clip,width=8cm]{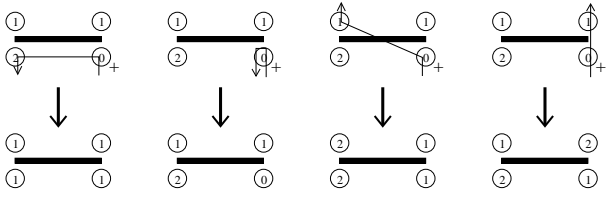}
\caption{The vertices (below) resulting from selecting the different exit legs on vertex 2 in Fig.~\ref{vertex2} (second from the left) when the entrance leg is leg 2 and the state change is in the opposite direction to that in Fig.~\ref{vertex2}.}
\label{vertex3}
\end{figure}

\begin{figure}
\includegraphics[clip,width=8cm]{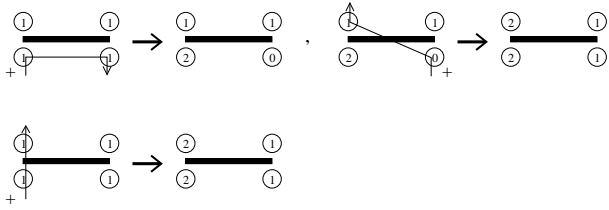}
\caption{Two different ways of arriving at vertex 3 in Fig.~\ref{vertex2}. In the top line the process goes from vertex 1 to 3 via vertex 2. While in the bottom line it goes directly from 1 to 3.}
\label{vertex4}
\end{figure}

Now start with the vertex $W_2$ and change the state on leg $2$ in the {\em opposite} way to what was done when leg $2$ was an exit leg. Exiting on any of the legs, $W_2$ has a similar decomposition as $W_1$:
\be
        W_2 = a_{21} + a_{22} + \cdots + a_{2n},
\ee
where now the entrance is on leg 2 on the vertex, with weight $W_2$, which differs from
vertex $1$ by having changed the states at leg $1$ and $2$. The weight 
$a_{21}$ corresponds to the process where the path enters at leg $2$ and 
exits at leg $1$. The states are changed in the {\em opposite} way to that 
when arriving at $W_2$ from $W_1$, and hence the process is undoing the 
changes and arriving back at $W_1$. In Fig.~\ref{vertex3} an example of this is shown.
From Eq.~(\ref{detbal1}) it follows 
that $a_{21} = a_{12}$. 
Now one can ask if exiting at leg 3 or 
higher yields the same vertex when starting from $W_2$ as it does starting 
from $W_1$. The answer to this is yes, because starting from $W_1$ one would 
change the state at legs $1$ and $3$ while starting from $W_2$ one would 
change the states at legs $2$ and $3$. But $W_2$ differs from $W_1$ only
by having different states at legs $1$ and $2$ and thus the state at leg 
$2$ is {\em changed twice} in opposite directions resulting in the same 
configuration $W_3$. An example illustrating this is shown in Fig.~\ref{vertex4}.
The weights are hence uniquely defined by this
procedure, and one is guaranteed that the only vertices which are related 
by the detailed balance equations are those which can be reached by changing
the state on the entrance leg together with the state on any exit leg of 
the reference vertex. The directed loop equations can therefore be written as
\be
\left(
\begin{array}{cccc}
        a_{11} & a_{12} & \cdots & a_{1n} \\
        a_{12} & a_{22} & \cdots & a_{2n} \\
        \vdots & \vdots & \vdots & \vdots \\
        a_{1n} & a_{2n} & \cdots & a_{nn}
\end{array}
\right)
\left(
\begin{array}{c}
        1 \\
        1 \\
        \vdots \\
        1 
\end{array}
\right)
=
\left(
\begin{array}{c}
        W_1 \\
        W_2 \\
        \vdots \\
        W_n 
\end{array}
\right) \label{dirloopeqs}
\ee 
where the matrix on the left hand side is a real symmetric $n \times n$ matrix
with all entries non-negative to avoid negative probabilities. The magnitudes
of the diagonal elements determine the probabilities for exiting
on the same leg as the entrance, the so-called bounce processes. The
bounce processes are generally ineffective as they do not change the spin
configuration and should be minimized.  
For a given model and parameters there are several such sets of equations.

Although the directed loop equations consist of $n=N_{\rm legs}$ equations this number is for many models in practice often reduced as one or more of the vertices arrived
at might not be allowed, that is they have zero weight. If so,
all entries in the matrix involving transitions to the disallowed vertex must be zero, which means that the corresponding column in the matrix is zero. The symmetry of the matrix implies that also the corresponding row is zero and so the dimensionality of the matrix is effectively reduced.
An example of this is the $s=1/2$ XXZ model where only three exit possibilities for each entrance leg is allowed, thus in this case the directed loop equation sets have dimensionality 3.
 
There is a general solution to the directed loop equations due to Sandvik\cite{loopsandvik} employed
in a number of works\cite{listworks}. This solution which we will label solution A reads
\be
a_{ij} = \f{W_i W_j}{W_1+W_2+\cdots +W_n},
\ee
which implies that the probabilities take the heat-bath form
\be
P(i \to j) = \f{a_{ij}}{W_i} = \f{W_j}{W_1+W_2+\cdots +W_n}.
\ee
The advantage of this solution is that it is general and easy to apply.
However it doesn't treat bounce processes different from other updates and
is thus expected not to be as effective in many cases. In Ref.~\cite{SS} it
was shown that there are other solutions which performs much better for the 
$s=1/2$ XXZ model.

The system of equations (\ref{dirloopeqs}) contains $n(n+1)/2$ unknowns and $n$ 
equations.
When $n \geq 2$ there are always more unknowns than equations; making many solutions possible. However if one seek solutions where
all bounces are absent, all diagonal elements zero, the number of
unknowns is reduced to $n(n-1)/2$ while the number of equations remains $n$.
So without bounces one finds that for $n=3$ the solution is unique while for $n > 3$ there are again many solutions. While this counting of unknowns and equations gives information about when one can expect solutions, it does not
ensure that the solutions are positive which is required in order to have positive probabilities. 

To investigate the solutions more closely let us start with the simplest case where the matrix is reduced to a $2 \times 2$-matrix.
\be
\left(
\begin{array}{cc}
        a_{11} & a_{12} \\
        a_{12} & a_{22} 
\end{array}
\right)
\left(
\begin{array}{c}
        1 \\
        1 
\end{array}
\right)
=
\left(
\begin{array}{c}
        W_1 \\
        W_2
\end{array}
\right) \label{2system}
\ee      
It is clear that a bounce-free solution $a_{11}=a_{22}=0$ can only occur
when $W_1=W_2$ for which $a_{12}=W_1$ implying that 
$P(1 \to 2) = P(2 \to 1)=1$.
When $W_1 \neq W_2$ bounces are necessary. There are several possibilities for writing down a solution in this case, but one solution which also generalizes to bigger matrices
is to choose to bounce off only the vertex with the biggest weight. If we assume $W_1 > W_2$ this solution is $a_{11}= W_1-W_2$, $a_{22}=0$ and $a_{12}=W_2$.
This gives $P(1 \to 2)=W_2/W_1$, $P(2 \to 1)=1$ and $P(1 \to 1)=1-W_2/W_1$.
Note that this bounce solution reduces to the bounce-free solution in the case $W_1=W_2$, thus whenever the directed loop equations reduces to a $2 \times 2$-system one can use the bounce solution above.

For a $3 \times 3$ system the bounce-free solution $(a_{11}=a_{22}=a_{33}=0)$ is unique and reads
\bea
	a_{12} & = & (W_1+W_2-W_3)/2 \nonumber \\
	a_{13} & = & (W_1-W_2+W_3)/2 \label{bouncefree3} \\
	a_{23} & = & (-W_1+W_2+W_3)/2 \nonumber
\eea
It is clear that this solution ceases to be a valid solution whenever
one of the weights is bigger than the sum of the two smaller weights.
In this case one needs again to include bounces. Again this can be done
as in the $2 \times 2$ case by only bouncing off the vertex with the biggest weight. Assuming that $W_1$ is the biggest weight this bounce
solution can be summarized as
\bea
   a_{11} & = & W_1-W_2-W_3, \nonumber \\
   a_{12} & = & W_2, \\
   a_{13} & = & W_3, \nonumber 
\eea
with all other $a$'s zero. This solution is complimentary
to the bounce-free solution Eq.~(\ref{bouncefree3}) being valid in the regime where the bounce-free solution is not valid. Furthermore these solutions are continuous at the boundary where $W_1=W_2+W_3$.
It is also interesting to see how these solutions reduces to the solutions described above for the $2 \times 2$-case when the smallest weight (assumed here to be $W_3$) goes to zero. Then clearly the bounce solution reduces to the $2 \times 2$-case. At first sight the bounce-free solution does not, however its regime of validity shrinks when $W_3 \to 0$ as $W_1 \geq W_2$ and
the bounce-free solution is only valid when $W_1 \leq W_2+W_3$. Thus the bounce-free solution is again only valid when $W_1=W_2$ and so also the bounce-free solution reduces to the solution found in the $2 \times 2$-case.

Going on to the $4 \times 4$-case it is clear that again we can write down the bounce solution by bouncing off only the vertex with the biggest weight.
In fact this solution can be generalized to any $n \times n$-matrix and
reads when we assume that $W_1$ is the biggest weight
\be
\left(
\begin{array}{cccc}
        a_{11} & W_2 & \cdots & W_n \\
        W_2 & 0 & \cdots & 0 \\
        \vdots & \vdots & \vdots & \vdots \\
        W_n & 0 & \cdots & 0
\end{array} \label{genbounce}
\right)
\ee
where $a_{11}=W_1-(W_2+W_3+\cdots+W_n)$.
This solution is valid when one weight is bigger than the sum of the
rest of the weights. This means in practice that bounces are only needed in
parameter regimes where one term in the Hamiltonian dominates.
The probability tables following from this solution has
a quite simple interpretation. The probability for moving between 
vertices other than the one with the biggest weight is zero
while that of moving from the largest weight
configuration to the smaller ones is the ratio of the smaller weight to the
larger weight and unity for the reverse process. The bounce probability is
unity minus the probabilities for moving to the smaller weight configurations.
 
Whenever the bounce solution 
above ceases to be valid one can write down a bounce-free solution. However, 
this is not unique as there are four equations with six unknowns. The
different solutions can be parametrized as follows
\bea
	a_{12} & = & (W_1+W_2-W_3-W_4)/2+a_{34} \nonumber \\
	a_{13} & = & (W_1-W_2+W_3-W_4)/2+a_{24} \label{parbouncefree} \\
	a_{23} & = & (-W_1+W_2+W_3+W_4)/2-\left( a_{34}+a_{24} \right) 
	\nonumber \\
	a_{14} & = & W_4 -\left( a_{34}+a_{24} \right) \nonumber
\eea
Here we have assumed that $W_1 \geq W_2 \geq W_3 \geq W_4$.
When $W_4 \to 0$ the requirement that $a_{14}$ be non-negative forces $a_{34}$ and $a_{24}$ both to zero. This solution with $a_{34}=a_{24}=0$ goes smoothly into the solutions for the $3 \times 3$ set described above in the limit $W_4 \to 0$. For later use we will term this solution where $a_{34}=a_{24}=0$ (for all $W_4$) for solution B. Solution B has the advantage that it is valid whenever the bounce solution above is not. It also goes continuously into the bounce solution at $W_1=W_2+W_3+W_4$. 

Solution B implies that the region where one can write down a solution
without bounces is given by 
\be
 -W_1+ W_2+W_3+W_4 \geq 0, \label{NoBounceCrit}
\ee 
where $W_1 \geq W_2 \geq W_3 \geq W_4$. While this was inferred from a special solution this result is in fact general. One can show on general grounds that the criterion allowing for a bounce-free solution is (for an $n \times n$-matrix)
\be
	-W_1+W_2+\ldots +W_n \geq 0.
	\label{bfallowed}
\ee

There are many bounce-free solutions when $n>4$. A general one
which reduces to the solution B above when $W_5=W_6=\ldots = W_n = 0$ is termed solution B1 here and is
\bea
	a_{12} & = & (W_1+W_2-W_3-W_4)/2 \nonumber \\
	a_{13} & = & (W_1-W_2+W_3-W_4)/2 \nonumber \\
	a_{23} & = & (-W_1+W_2+W_3+W_4)/2 \nonumber \\
	a_{14} & = & W_4 - W_5/2 \nonumber \\
	a_{15} & = & (W_5-W_6)/2  \nonumber \\
	       & \vdots & \\
	a_{1,n-1} & = & (W_{n-1} - W_{n})/2 \nonumber \\
	a_{1n} & = & W_n/2 \nonumber \\
	a_{45} & = & W_5/2 \nonumber \\
	a_{56} & = & W_6/2 \nonumber \\
	       & \vdots & \nonumber \\
	a_{n-1,n} & = & W_n/2 \nonumber
\eea
The requirement that B1 reduces to solution B above restricts
$-W_1 +W_2 +W_3 + W_4 \geq 0$.
  
The above matrix framework also holds in the case where there is no conservation law which dictates the state change on the exit leg once the state change on the entrance leg is given. Then the matrix dimension is increased to $n=N_{\rm legs} \times N_{\rm states}$, where $N_{\rm legs}$ is the number of legs on the vertex and $N_{\rm states}$ are the number of allowed state changes on each exit leg. This is because one must here also take into account all possible state changes on the exit legs. This includes the possibility where the state on the exit leg remains unchanged, which as described on the previous section will terminate the path.
The numbers $1$ to $n$ thus each have a vertex type, a leg and an update type associated with them. Here also the state change of the exit leg $j$ in the process $a_{ij}$ is opposite to the state change on the entrance leg $j$ in the process $a_{jk}$.

\section{Spin-s XXZ-models}
Before we study the efficiency of different solutions let us consider an example.
Consider the spin-$s$ XXZ model with nearest neighbor interactions.
We have also added a magnetic field, an interaction proportional to
$(S^z)^2$ and a physically  unimportant constant $C$
\bea
{\cal H} & = & -\sum_{<ij>} \left\{ S^x_i S^x_j + S^y_i S^y_j
                           - J_z S^z_i S^z_j +C\right\} \\
	 &  & +\sum_i \left\{ - h S^z_i +v S^{z2}_i \right\} 
			   \nonumber 
\eea
We take the exchange coupling to be ferromagnetic and choose units such that its magnitude is unity.
Using the normalization of the ladder operators $S_{\pm} = S_x \pm i S_y$
\be
   S_{\pm} |s,m \rangle = \sqrt{(s \mp m)(s \pm m+1)} |s,m \pm 1 \rangle
\ee
where $m \in \left[-s,s \right]$ is the $S^z$ quantum number which labels the state on a 
vertex-leg, 
it is easy to see that the different vertex weights for this model are
\begin{widetext}
\bea
  W\left( n \pm 1, m \mp 1, n, m \right) & = & \f{1}{2}
  \sqrt{(s\mp n)(s\pm n+1)(s\pm m)(s \mp m+1)} \nonumber \\
  W\left( n , m, n, m \right) & = & C-\left( J_z n m - \bond{h} \left( n+m \right)
  + \bond{v} \left( n^2 +m^2 \right) \right), \nonumber 
\eea
\end{widetext}
where the arguments of $W$ represent the states on legs $1,2,3$ and $4$ labeled as in Fig.~\ref{vertex1}, and $\bond{a} = a/Z$, where $Z$ is the coordination number of the lattice.
$C$ must be chosen
such that all diagonal weights are positive. Generally we will choose $C$ to be slightly above its minimal value. The antiferromagnetic case can be studied on bipartite lattices where the spins on one of the sublattices are rotated an angle $\pi$ about the $z$-axis giving a sign change to the spin-flip-term but not to the $J_z$ term.

\begin{figure}
\includegraphics[clip,width=8cm]{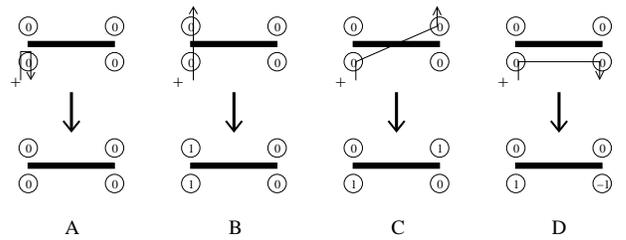}
\caption{Update of a vertex for the spin-1 XXZ model which results in
the 4 allowed vertices shown below. Here we have labeled the single-site 
states with the eigenstates of $S^Z$: $m \in [-1,0,1]$. From left to right the resulting vertex weights are $C$, $C+\bond{h}-\bond{v}$, $1$ and $1$. If instead one had considered the update where the entrance spin is decreased by one unit, the spins on the resulting vertices would have the opposite sign and the only change in the weights would be that the second vertex from the left would have weight $C-\bond{h}-\bond{v}$.}
\label{spin1upd}
\end{figure}

For general $s$ the maximal dimensionality  of the directed loop equation sets for this model is 4. However, they do not all have dimension 4.  For the vertices containing a spin with the maximum ($m=s$) spin the equation set for updates which attempts to increase this spin further has dimension 3 as this attempt will lead to a vertex with zero weight.
This holds also for updates attempting to decrease a spin below its minimum value. It thus follows that all sets of directed loop equations for the $s=1/2$ XXZ model has at most dimension $3$\cite{SS}. 
For $s=1$ the update of the vertex shown in Fig.~\ref{spin1upd} is described by an equation set with dimensionality 4. This update together with the update where the spin on the same vertex is decreased, their symmetry related updates (entering from another in-leg) and their reverse updates(going backwards) are in fact the only updates for the $s=1$ model governed by an equation set with dimension 4. The other equation sets have dimensions 2 and 3. 

\begin{figure}
\includegraphics[clip,width=8cm]{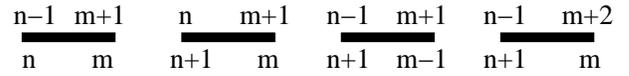}
\caption{The vertices resulting from increasing the spin on the lower left leg of the off-diagonal vertex, shown left. The right two vertices are not allowed in the XXZ model, and so the directed loop equation set reduces to having dimensionality 2. The circles around the leg states are omitted for clarity.
\label{oppvertex}}
\end{figure}

It is interesting to ask for the region in parameter space where the ineffective bounce processes can be avoided. The defining feature of this region is that Eq.~(\ref{NoBounceCrit}) should hold for all equation sets in the model. For $s=1/2$ this region was shown in Ref.~\cite{SS} to be the the region defined by $|J_z|+2|\bond{h}| \leq 1$. That is for XY-like anisotropies and moderate fields. To generalize this region to arbitrary spin-$s$ one can consider increasing the state on the lower left leg by unity on a diagonal vertex with weight $W(n,m,n,m)$. The directed loop equations then relates this vertex to the vertices $W(n+1,m,n+1,m)$, $W(n+1,m-1,n,m)$ and $W(n+1,m,n,m+1)$. In the case $m=0$ the off-diagonal weights are equal, while for $m=\pm s$ one of the off-diagonal weights vanishes.
Now choose $C$ such that the weights of the diagonal vertices are always larger than the off-diagonal ones. While this is a matter of choice in the SSE method it is always true in the Loop algorithm as the diagonal vertices are of order unity while the off-diagonal ones are of order the Trotter spacing.
Then the inequality (\ref{NoBounceCrit}) takes the form
\bea
| J_z m-\bond{h} +\bond{v}(2n+1) |
& \leq & \f{1}{2} \sqrt{(s+n+1)(s-n)} \nonumber \\
&      & \times \left( \sqrt{(s-m+1)(s+m)} \right. \nonumber \\
&      & + \left. \sqrt{(s+m+1)(s-m)} \right) \nonumber \\
\eea
where $n \in [-s,s-1]$ and $m \in [-s,s]$.
The most restrictive case is for $n=s-1$ and $m=s$ for which the bounce-free criterion is
\be
|J_z|s+|\bond{h}|+|\bond{v}|(2s-1) \leq s. \label{spinsbouncefree}
\ee
The same inequality is obtained by considering lowering a spin on a diagonal vertex and when a spin on an off-diagonal vertex is changed in the same direction as it is being changed by the off-diagonal operator. However when a spin on an off-diagonal vertex is changed in the direction {\em opposite} to how it is changed by the operator a stricter condition is obtained, see Fig.~\ref{oppvertex}. This is because there are no operators changing the spin on a site by two units in the XXZ-model. To avoid bounces in this case 
\be
	\left( s+m \right)\left( s-m+1 \right) = 
	\left( s+m+1 \right)\left( s-m \right) \label{oppvcond}
\ee
where $-s+1 \leq m \leq s-1$. Thus this condition does not apply for $s=1/2$. The condition Eq.~(\ref{oppvcond}) is equivalent to $m=-m$, which only is satisfied for $m=0$. Thus for $s \leq 1$ the condition (\ref{oppvcond}) does not constrain the no-bounce parameter region. However for $s>1$ one always need bounces for these kind of vertices, and so for $s>1$ it is not possible to find an algorithm within the framework presented here which is entirely without bounces\cite{Alet}.

For $s > 1/2$ it is also possible to consider updates that change the quantum numbers by more than one unit. However because there are no non-zero off-diagonal terms where the magnetization at a site is changed by two units such an
update will be described by an equation set of dimensionality 2. 
Thus for almost all parameters will this update
contain bounces and because the path without bounces is deterministic (it is determined by the straight-through process for diagonal vertices and the diagonal process for off-diagonal ones) there will be a sizable probability for the process to bounce back and forth along the predetermined path before it possibly ends by retracing its path all the way to the starting point without having done any changes. It is thus not expected that inclusion of these updates will make the simulation more efficient.

\begin{figure}
\includegraphics[clip,width=8cm]{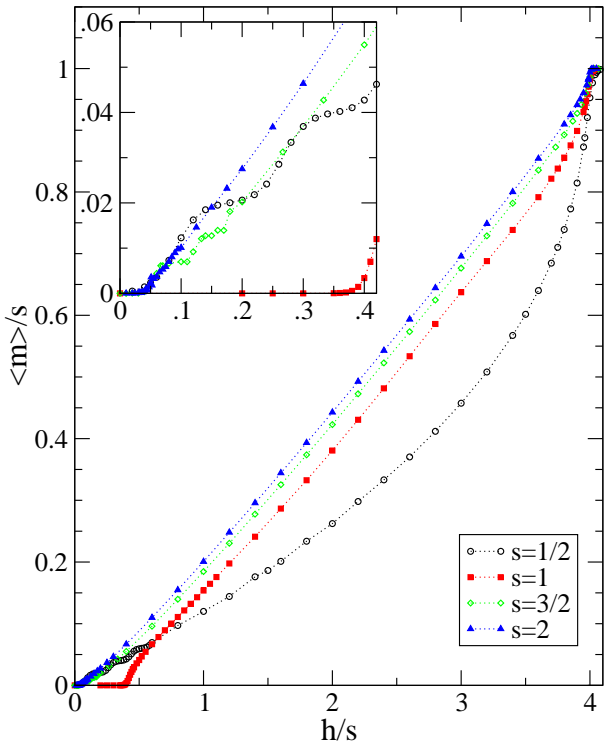}
\caption{Magnetization per site and spin magnitude as function of h/s for 100 sites Heisenberg antiferromagnetic chains having different spins $(\beta = 100)$. The inset shows the behavior at low fields. The dotted lines are guides to the eye.}
\label{spinmagn}
\end{figure}

To demonstrate that different spin magnitudes can be simulated efficiently using the same basic code with changing just the number of different states on a site and the matrix elements of the Hamiltonian we show in   
Fig.~\ref{spinmagn} the magnetization curves for different 100 sites Heisenberg antiferromagnetic spin-chains with $s=1/2$, $1$, $3/2$ and $2$ at inverse temperature $\beta J = 100$. At low fields one can clearly see that the integer spin chains have a gap while the half-integer ones are gap-less. At low fields the stair-case finite size effects are also clearly seen for the half-integer chains. 
A typical point on the $s=1/2$ curve is based on an average of 10 bins each with $10^4$ MCS ($5 \times 10^4$ MCS for equilibration) and took about 20 minutes on a single processor 868 MHz Intel Pentium III. For comparison a typical point on the $s=1$, $s=3/2$ and $s=2$ curves using the same number of equilibration and measurement steps took about $2$, $5$ and $8$ hours respectively. The $s=1/2$,
$1$, $3/2$ and $2$ simulations involves $6$, $17$, $34$ and $57$ non-zero vertices respectively.

\section{Efficiency}
In Ref.~\cite{SS} a number of examples for the $s=1/2$ XXZ model showed that algorithms minimizing the number of bounces is generally more effective than those where no such minimization is attempted. 

In the $s=1/2$ XXZ-model the directed loop equations have dimensions $3$. Thus in the regime where a bounce-free solution exists, it is unique. However, when the directed loop equations have dimension $4$ or greater the bounce-free solution is not unique. The solutions to the bounce-free 4-dimensional case can be parametrized as shown in Eqs.~(\ref{parbouncefree}).
We will now investigate how to choose the most efficient of these.
This cannot be done in complete generality as a general physical model involves many
equation sets relating the different vertices, and as we will see the overall efficiency depends on how the sets are interconnected. 
However we can pick a specific model and parameters, and try out different solutions there. For this we pick
a simple model with a 4-dimensional equation set, the $s=1$ Heisenberg model, and measure autocorrelations for the staggered magnetization as a function of different bounce-free solutions.       

We use the integrated autocorrelation time as a measure of efficiency\cite{evertzchapter}. It is a measure for how many subsequent Monte Carlo steps (MCS) are needed in order to obtain statistically independent configurations. A low value indicates an efficient algorithm.
The integrated autocorrelation time for a quantity ${\cal O}$ is defined as
\begin{equation}
\tau_{\rm int}[{\cal O}] = {1\over 2} + \sum_{t=1}^\infty A_{\cal O}(t),
\end{equation}  
where $A_{\cal O}(t)$ is
\begin{equation}
A_{\cal O}(t) = {\langle {\cal O}(i+t){\cal O}(i) \rangle - \langle {\cal O}(i) \rangle ^2 \over 
\langle {\cal O}(i)^2 \rangle - \langle {\cal O}(i) \rangle^2 },
\label{atau}
\end{equation}
where $i$ and $t$ are Monte Carlo times, measured in units of one MCS. A MCS is defined here by adjusting the number of loop updates during equilibration such that every vertex in the 
linked list is on average visited twice (not counting bounces) in one MCS. 

There are only two inequivalent 4-dimensional equation sets for the $s=1$ Heisenberg model. See Fig.~\ref{spin1upd} for the vertices belonging to one of these sets. The vertices belonging to the other set are labeled A$^\prime$, B$^\prime$, C$^\prime$ and D$^\prime$, and are obtained by changing the sign on all states in the corresponding unprimed vertices in Fig.\ref{spin1upd}.  

In the zero field Heisenberg case ($J_z=1$, $h=v=0$) all weights of the vertices shown in Fig.~\ref{spin1upd} can be chosen to be equal ($C=1$).
 When two or more weights are equal there is an ambiguity in the ordering of vertices when ordered according to their size as is assumed in the specification of the solution in Eqs.~(\ref{parbouncefree}). However, a specific
ordering of the vertices and values of $a_{24}$ and $a_{34}$
can be shown to be equivalent to a solution for a different ordering of the equal-weights vertices with other values of $a_{24}$ and $a_{34}$. The explicit relations are shown in Appendix~\ref{Appendix}.  Therefore it suffices to choose one ordering which we here choose to be ABCD, meaning $W_1=W_A$, $W_2=W_B$, $W_3=W_C$ and $W_4=W_D$ where the letters refer to the letters in Fig.~\ref{spin1upd}, and investigate efficiency, as measured by the integrated autocorrelation function for the staggered magnetization, as functions of $a_{24}$ and $a_{34}$. Because of the time-reversal symmetry we choose the ordering A$^\prime$B$^\prime$C$^\prime$D$^\prime$ for the other set and use equal values of $a_{24}$ and $a_{34}$ for the two sets.

To avoid searching the whole two-dimensional space of these two quantities we will restrict ourselves to three lines where $a_{24}=a_{34}=a/2$, $a_{24}=a$, $a_{34}=0$ and $a_{24}=0$, $a_{34}=a$ respectively. The quantity $a$ is restricted to the set $[0,W_\Delta]$ where $W_\Delta= (-W_1+W_2+W_3+W_4)/2$. This restriction follows from the requirement of having non-negative weights.

\begin{figure}
\includegraphics[clip,width=8cm]{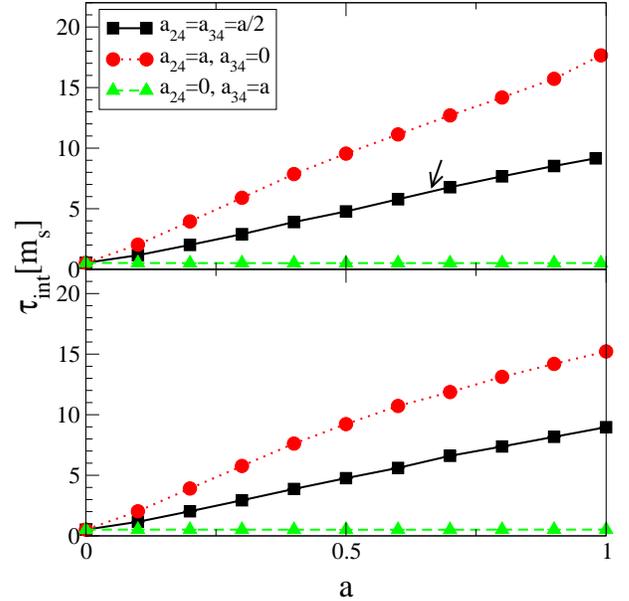}
\caption{Integrated autocorrelation times for the staggered magnetization of a 64-site Heisenberg ($J_z=1$) spin-1 chain at $\beta =16$ as functions of different values of $a_{24}$ and $a_{34}$. The order of the vertices when having equal weights is ABCD. Solid lines indicate the parametrization $a_{24}=a_{34}=a/2$, dashed lines: $a_{24}=0$, $a_{34}=a$ and dotted lines are $a_{24}=a$, $a_{34}=0$. The top panel is for the case where all weights are equal, $C=1$, $h=0$. The bottom panel is for $C=1.5$, $h=0$. The arrow indicates the location of the fully symmetric algorithm where $a_{ij}=1/3$ for $i \neq j$.}
\label{auto1}
\end{figure}

In Fig.~\ref{auto1} we show integrated autocorrelation times for the staggered magnetization for values of $a_{24}$ and $a_{34}$ along
the three lines specified above.
In the top panel all weights in both 4-dimensional equation sets are equal and it is clearly seen that having $a_{24}=0$ gives the lowest autocorrelation times, being close to the minimal value 0.5 in the whole range of $a_{34}$. Naively one would think that for equal weights the most effective solution should be the most symmetric one: $a_{ij}=1/3$, for all $i \neq j$. 
The arrow in the top panel shows the autocorrelation time for this case. 
This is clearly {\em not} the most efficient solution. 
One should keep in mind that the effectiveness as measured here depends in general not just on the rules for the directed-loop equation sets seen individually, but also on how different sets are are interconnected. The asymmetry observed above should probably be attributed to the diagonal vertex with an up and a down spin which has double the weight of the other vertices. Although it is not a part of the two 4-dimensional equation sets it still plays a role making an asymmetry in how to choose the most effective solution.    

Letting $C>1$ all four weights are no longer equal. However the two diagonal ones are still equal to each other as well as are the two off-diagonal ones. Fig.~\ref{auto1} bottom panel shows autocorrelation times when $C=1.5$ still with the same ordering of vertices as in the top panel. Again we see that $a_{24}=0$ is favorable for the efficiency. In Ref.~\cite{sara} the spin-1 Heisenberg model with disorder was studied using directed loops using the solution where all diagonal updates were excluded and other processes have equal probabilities. In the language used here their solution corresponds to the case where $a_{24}=0$ and $a_{34}=1/2$ with the orderings ABCD and A$^\prime$B$^\prime$C$^\prime$D$^\prime$.

\begin{figure}
\includegraphics[clip,width=8cm]{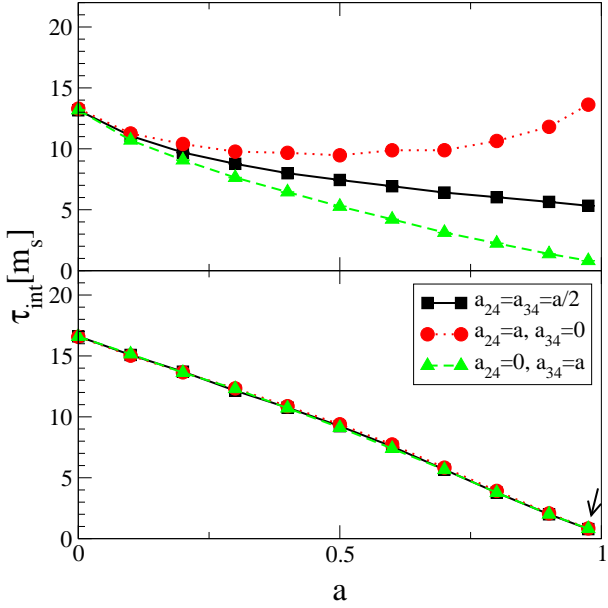}
\caption{Same as Fig.~\ref{auto1} except that $h=0.1$. In the upper panel the ordering of vertices is $BACD$ and A$^\prime$B$^\prime$C$^\prime$D$^\prime$ while
it is $BACD$ and A$^\prime$B$^\prime$D$^\prime$C$^\prime$ in the lower panel.
The arrow indicates the best algorithm found in Ref.~\cite{HK}.
\label{auto2} }
\end{figure}

In a magnetic field the time-reversal symmetry is broken and the two 4-dimensional equation sets are no longer equal.
Furthermore the diagonal vertices in each of these sets are not equal and it suffices to choose the ordering of the two off-diagonal ones. 
Fig.~\ref{auto2} shows
integrated autocorrelations for the staggered magnetization at $h=0.1$ for different values of $a_{24}$ and $a_{34}$. 
In the upper panel the ordering of the vertices are $BACD$ and A$^\prime$B$^\prime$C$^\prime$D$^\prime$ while it is $BACD$ and A$^\prime$B$^\prime$D$^\prime$C$^\prime$ in the lower panel. In both panels we have used the same values of $a_{24}$ and $a_{34}$ for the two sets. The most efficient algorithm using the ordering in the upper panel is found for $a_{24}=0$ and $a_{34}=W_\Delta$. For the ordering used in the lower panel, the efficiency is seen to depend solely on the sum $a_{24}+a_{34}$ and is maximized for $a_{24}+a_{34}=W_\Delta$.
A similar analysis was carried out by Harada and Kawashima\cite{HK}. In the language used here their parametrization corresponds to a parametrization of the ordering used in the lower panel where $a_{34}=0$, and $a_{24}$ was varied. The algorithm performing best at a field $h=0.1$, their solution 1, corresponds to the case $a_{24}=W_\Delta$ which indeed also is among the most efficient algorithms found here.

It is interesting to compare these autocorrelation times with those obtained using solution A. The integrated autocorrelation times for the staggered magnetization using solution A corresponding to the top, bottom panels of Fig.\ref{auto1} and to Fig.\ref{auto2} are $11.6$, $12.5$ and $12.0$ respectively, which is significantly more than the most efficient algorithms which have autocorrelation times close to $0.5$. 

For higher magnetic fields the variation of autocorrelation times with the parameter $a$ is similar to what is shown for $h=0.1$ in Fig.~\ref{auto2}. However at high fields the differences between the least and most efficient choices are less pronounced than in small fields. 
  
\section{Interacting bosons}
The directed loops can also be applied to the boson Hamiltonian
\bea
	H & = & \sum_{<ij>} \left\{ 
	-t \left( a^\dagger_i a_j + \rm{h.c.} \right)
	+v n_i n_j -C \right\}
	\nonumber \\ 
	& + & \sum_{i} \left\{ - \mu n_i + \f{U}{2} n_i (n_i-1) 
	               \right\} \label{bosHamiltonian}
\eea
which describes interacting bosons hopping on a lattice. 
$a_i^\dagger$, $a_i$ and $n_i$ are the boson creation, annihilation and
density operator on site $i$ respectively.
The single-site states are labeled by the boson occupation number. The
vertex weights are
\bea
  W(n,n^\prime,n + 1,n^\prime - 1) & = & 
  t \sqrt{(n+1)n^\prime} \left( 1-\delta_{n,n_x} \right) \nonumber \\
  W(n,n^\prime,n - 1,n^\prime + 1) & = & t \sqrt{n(n^\prime+1)}
   \left( 1-\delta_{n^\prime,n_x} \right) 
                            \label{bosweights}\\
  W(n,n^\prime,n,n^\prime) & = & C-\left[ v n n^\prime - \bond{\mu} \left( n+n^\prime \right) \right. \nonumber \\
    &  & \left. +\f{\bond{U}}{2} \left( n(n-1) +n^{\prime}(n^\prime-1) \right) \right], \nonumber 
\eea
where $\bond{a}=a/Z$, and 
$n,n^\prime \in [ 0,n_x ]$. 
$C$ must be chosen such that the diagonal vertices are non-negative.
However, this can not always be achieved as in principle
the boson occupation number can be arbitrarily large. So in practice
when $\bond{U} > -|v|$ one must put an upper limit, $n_{x}$, 
to how many bosons can occupy a site.
The $(1-\delta_{n,n_x})$ factors are needed to
implement this. They prevent the boson occupation from increasing above $n_x$.
Imposing this restriction the boson system has as many states on a single
site as a spin-s model where $2s+1 = n_x+1$. Furthermore 
there is a one-to-one correspondence between the non-zero vertices
in the boson model and the non-zero vertices in the spin-s XXZ model,
although their values are of course different. 
Thus the only change needed to the spin-s program code is to change the numerical
values for the weights. 
A special example of this is the well-known mapping in the hard-core limit $(U/t \to \infty)$ where 
one can set $n_x=1$ without penalty and thus the boson model is mapped 
onto a spin-1/2 model. The error introduced by imposing a restriction on
the boson occupation number can be controlled by repeating runs with
different $n_x$'s. One expects only small variations in the results as long as $n_x$ is bigger than the typical occupation number on each site. In the simulations described below we use $n_x=4$ for which there are $57$ non-zero vertices. 
  
\begin{figure}
\includegraphics[clip,width=8cm]{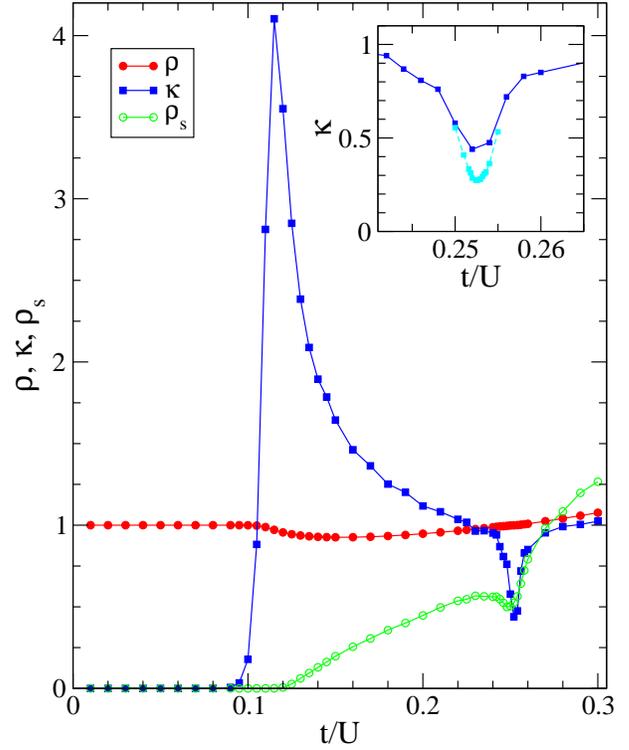}
\caption{Density $\rho$, compressibility $\kappa= \partial \rho/\partial \mu$ and superfluid density $\rho_s$ for a $N=128$ site 1D boson Hubbard model with $U=1$, $\mu/U=0.17$ as functions of the hopping $t/U$. The inverse temperature is $\beta U = 320$. The inset shows the compressibility in the region of the reentrant insulating phase. The lower curve in the inset is for $N=256$ and $\beta U=480$. The superfluid density shown is measured as the square of the boson world-lines spatial winding number\cite{Ceperley}.
\label{monien}}
\end{figure}

\begin{figure}
\includegraphics[clip,width=8cm]{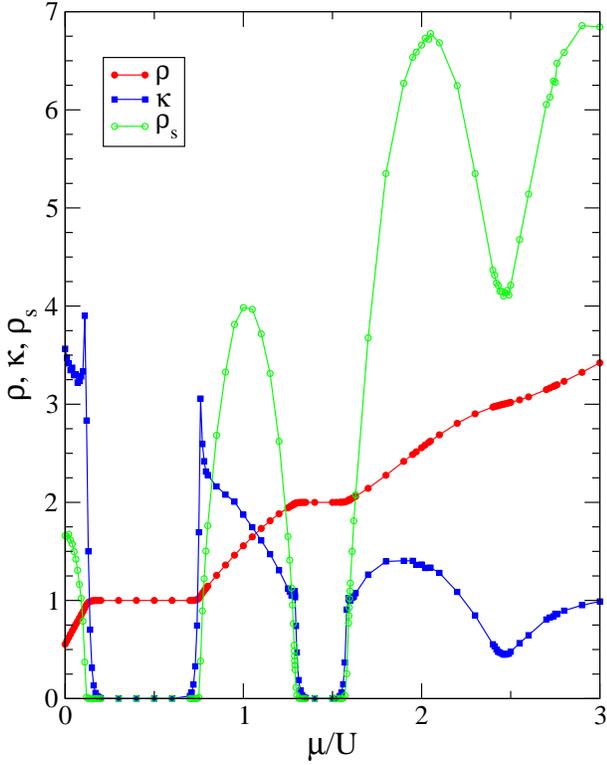}
\caption{Density $\rho$, compressibility $\kappa= \partial \rho/\partial \mu$ and superfluid density $\rho_s$ for the $32 \times 32$ boson Hubbard model with $t/U=0.03$ and $\beta t = 2.5$ as functions of the chemical potential $\mu/U$. 
\label{bosmu}}
\end{figure}

\begin{figure}
\includegraphics[clip,width=8cm]{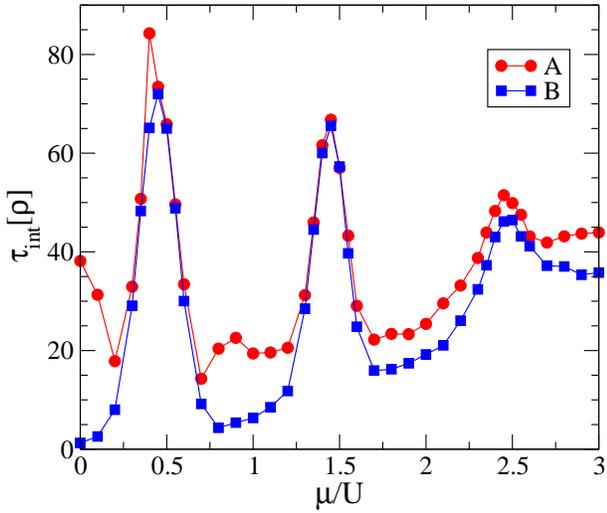}
\caption{Integrated autocorrelation times for the density $\rho$ in a $12 \times 12$-Bose-Hubbard model. $U/t = 33.333$ and $\beta t=0.5$ for the two solutions A and B.
\label{bosauto}}
\end{figure}

To show the efficiency of solution B applied to this boson system we show in Fig.~\ref{monien} the reentrant behavior of the Mott insulating phase in 1D first predicted from a DMRG calculation\cite{kuhner}. 
In Fig.~\ref{bosmu} we show how the density $\rho$, compressibility $\kappa=\partial \rho/\partial \mu = N\beta( \langle \rho^2 \rangle - \langle \rho \rangle^2)$ and superfluid density $\rho_s$  varies for a 2D Bose-Hubbard model as the chemical potential is increased keeping $t/U$ fixed. One can clearly see the insulating regions as well as the singularities in the compressibility at the first and second superfluid-insulator transition. Because of the relatively big value of $t/U$, being near the tip of the lowest Mott lobes, the insulating regions become rapidly narrower, and the gaps are relatively small making it necessary to lower the temperature further to see the singularities at the higher-$\mu$ superfluid-insulator transitions.
Fig.~\ref{bosauto} compares integrated autocorrelation times for the density as measured using solution A and B for a smaller but similar system to that shown in Fig.~\ref{bosmu}. The differences in autocorrelation times are most pronounced
at low values of the chemical potential and away from the peaks seen in the 
Mott insulating regions which arises from the small denominator in Eq.~(\ref{atau}).   

To find the regime where it is possible to use bounce-free algorithms for this Hamiltonian we consider
loops changing the number of bosons by unity. 
The conservation law here is that the Hamiltonian conserves the total number of bosons, $n_1+n_2=n_3+n_4$. 
\begin{figure}
\includegraphics[clip,width=8cm]{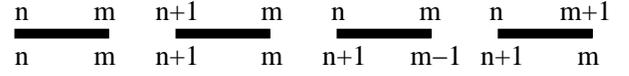}
\caption{Increasing the occupation on the lower left leg results in the following four vertices.(The circles on the legs have been omitted.)}
\label{vertex5}
\end{figure}    
Now increase the number of bosons on the lower left leg of a diagonal vertex by unity, then the vertices related
by the directed loop equations are shown in Fig.\ref{vertex5} and their weights are from left to right 
\bea
        W_1 & = & W(n,m,n,m) \nonumber \\
	W_2 & = & W(n+1,m,n+1,m)  \\
	W_3 & = & W(n+1,m-1,n,m) \nonumber \\
	W_4 & = & W(n+1,m,n,m+1) \nonumber 
\eea
where $W(n_1,n_2,n_3,n_4)$ refers to Eq.~(\ref{bosweights}), and
$n \in [ 0,n_x-1]$, $m \in [0,n_x]$. The difference between the diagonal weights is $W_1-W_2 = vm+\bond{U}n-\bond{\mu}$.
Now choose $C$ such that the biggest diagonal weight ($W_1$ or $W_2$) is always bigger than any of the off-diagonal weights $W_3$ and $W_4$. One then finds
from Eq.~(\ref{NoBounceCrit}) that in order to avoid bounces
\be
| v m+\bond{U}n-\bond{\mu} | \leq 
t \sqrt{n+1} \left( \sqrt{m} + \sqrt{m+1} \left( 1-\delta_{m,n_x} \right) \right)
\ee
where $n \in [0,n_x-1]$, $m \in [0,n_x]$. 
The same inequality is also obtained when considering lowering the boson occupation value on a diagonal vertex. In the hard-core case where $n_x=1$ this criterion reduces to $|\bond{\mu}| \leq t$ and $|v-\bond{\mu}| \leq t$. 
Changing the boson occupation on a leg on an off-diagonal vertex in the opposite direction to how the state is changed by the operator results in an additional criterion as was the case for spin-s models. Here this is
\be
     t \sqrt{n} = t\sqrt{n+1}
\ee
for $n_x-1 \geq n \geq 1$. Thus this does not apply in the hard-core boson case. For $n_x >1$ the equality is never satisfied, and so bounces are always needed for these kind of vertices when $n_x > 1$.  

\section{1D Fermions with spin}
The rules described here applies also to 1D spinful fermions\cite{SandvikJPA}. However for (anti)periodic boundary conditions the simulation is restricted to a system where the number of spin-ups and downs are both an (even)odd number.
Thus with these boundary conditions one must carry out the simulation at so low temperatures that only configurations having a fixed (even)odd number of up and down spins contributes. 
For open boundary conditions there are no such restrictions. 

Consider the 1D fermion Hamiltonian
\bea
    H & = & \sum_{<ij>} \left\{ -t \left( c^\dagger_{i\sigma} c_{j \sigma} + \rm{h.c.} \right) 
            +V (n_i-1) (n_j-1) 
            \right. \nonumber \\ 
      &   & \left. -J_\perp \left( S^x_i S^x_j + S^y_i S^y_j \right) + J_{z} S^z_i S^z_j -C \right\} \\
      &   & +\sum_{i} \left\{ U \left(n_{i \up}-\f{1}{2} \right) 
	                        \left(n_{i \down}-\f{1}{2} \right)  
				- \mu n_i -H_z S^z_i \right\}
     \nonumber
  \eea
where $c^\dagger_{i\sigma}$, $c_{i\sigma}$ and $n_{i \sigma}$ are the creation, annihilation and density operators of a fermion with spin $\sigma$ on site $i$. $n_i = n_{\up i} + n_{\down i}$ and $\vec{S}_i = c^\dagger_i \vec{\sigma} c_i/2$ are the particle density and spin operator on site $i$. The states on each site are labeled by the charge and spin $|q,s \rangle$ so that $|0 \rangle \equiv |0,0 \rangle$,$|\up \rangle \equiv |1,\f{1}{2} \rangle$, 
$|\down \rangle \equiv |1,-\f{1}{2} \rangle$ and $|\up \down \rangle \equiv |2,0\rangle$.
The weights of the diagonal vertices can be read directly off the Hamiltonian above and are
\begin{widetext}
\bea
W(q_1s_1,q_2s_2,q_1s_1,q_2s_2) & = &
 C -\left\{ \f{^{}}{} J_z s_1 s_2 + V(q_1-1)(q_2-1) \right. \label{fermionHamiltonian}\\
   &  & \left.  +\f{\bond{U}}{2}\left[ (q_1-1)^2+(q_2-1)^2-1 \right]
   - \bond{H_z}\left( s_1+s_2 \right)-\bond{\mu} \left( q_1+q_2 \right)        \right\},  \nonumber 
\eea
\end{widetext}
where again $\bond{a} = a/Z$ and $Z$ is the coordination number. 
As the states will be represented in terms of occupation number states, which are bosonic states, one needs a convention for how the phases are related to the
fermionic states. We follow ref.~\cite{SandvikJPA} and define
\bea
 c^\dagger_{i\up} | .... n_{i\up}=0 ...\rangle & = & (-1)^{n_{i\up}^<}
                                     | .... n_{i,\up}=1 ...\rangle \\
 c^\dagger_{i\down} | .... n_{i\down}=0 ...\rangle & = & (-1)^{n_\up+n_{i\down}^<}
                                     | .... n_{i,\down}=1 ...\rangle
\eea
where $n_{i\up}^<$ is the operator counting the number of particles with spin
up on sites less than $i$, and $n_\up$ is the total number operator for spin-up
particles. A similar definition holds for the annihilation operators.
The explicit $n_\up$ for the spin down states is to ensure  anticommutation between fermi operators with different spin indeces.
With (anti)periodic boundary conditions 
\bea
   c^\dagger_{N+1,\up} & = & (-1)^P c^\dagger_{1,\up} \\
   c^\dagger_{N+1,\down} & = & (-1)^P c^\dagger_{1,\down}
\eea
where $P=(-1)0$ it follows that
for the off-diagonal weights to be non-negative $n_\down$, $n_\up$ must both be (even)odd.

The weight of all off-diagonal vertices where one particle, and only one, is transferred from one site to the other is $t$. The off-diagonal vertex where two-particles are interchanged, the spin-flip vertex, has weight $W(_{\up \down}^{\down \up})=J_\perp/2$. In all there are $34$ allowed vertices ($32$ if $J_\perp=0$).  
We consider four types of updates: Adding/removing a spin-up/down particle.
The conservation law here is the conservation of charge and spin. That
is: $q_1+q_2=q_3+q_4$ and $s_1+s_2=s_3+s_4$.

This conservation law is in fact so strong that no 4-vertex relations exists. 
This can be seen by considering the process of adding a spin-up particle
to the lower left leg on a diagonal vertex. In this case the update where the exit is on the upper left leg, the ``continue straight through'', will be allowed as well as the bounce. In addition there is one and only one off-diagonal vertex that arises: If the right legs contains an up-spin, then the lower right leg is the allowed exit. The upper right leg is the allowed exit if the right legs does not contain an up spin. Thus only three vertices are related by the directed loop equations. It is quite clear that this also holds for all the other update processes on a diagonal vertex. One must also consider adding a spin-up particle to the off-diagonal spin-flip vertex. This results again in three allowed vertices, but now they are all off-diagonal, two with weight $t$ and one with weight $J_\perp/2$. Finally adding a spin-up particle to a leg on an off-diagonal vertex with weight $t$ results in one of the two above mentioned equation sets or in a 2-dimensional equation set with equal weights. Thus for this model all the directed loop equations have dimensions $3$ or less. 

To find the regime where bounces can be avoided we employ Eq.~(\ref{NoBounceCrit}) with $W_4=0$. Choosing $C$ big enough so that one of the diagonal vertices always have the biggest weight it follows
that 
\be
t \geq |J_z|/4+|V| +|\bond{U}|/2+|\bond{H_z}|/2+|\bond{\mu}|
\ee
in order to avoid bounce-solutions to the equation sets where two diagonal and one off-diagonal vertex are related. $C$ cancels as it occurs on both sides of the inequality.
For the set where three off-diagonal vertices are related one must have
\be
t \geq J_\perp /4
\ee
to avoid bounces if $t \leq J_\perp/2$, otherwise it suffices that $J_\perp$ is non-negative. 

To compare the efficiency of solution B with solution A
we have measured the integrated autocorrelation times for the charge density wave (CDW) order parameter at $q=\pi$ defined as
\bea
     {\cal O}_{\rm CDW}(q) & = & \langle \rho_c(q) \rho_c(-q) \rangle 
\eea
where
\bea
 \rho_c(q) & = & (1/N) \sum_r e^{iqr} (n_{\up r} + n_{\down r})
\eea
as functions of $V$ for $U/t=2$, $N=16$ and $\beta t=32$ for both
solution $A$ and $B$. 
While the autocorrelation times are quite long in this case considering the relatively small system size the solution B is more efficient than solution A. The difference in autocorrelation times are largest for small values of $V$.    
\begin{figure}
\includegraphics[clip,width=8cm]{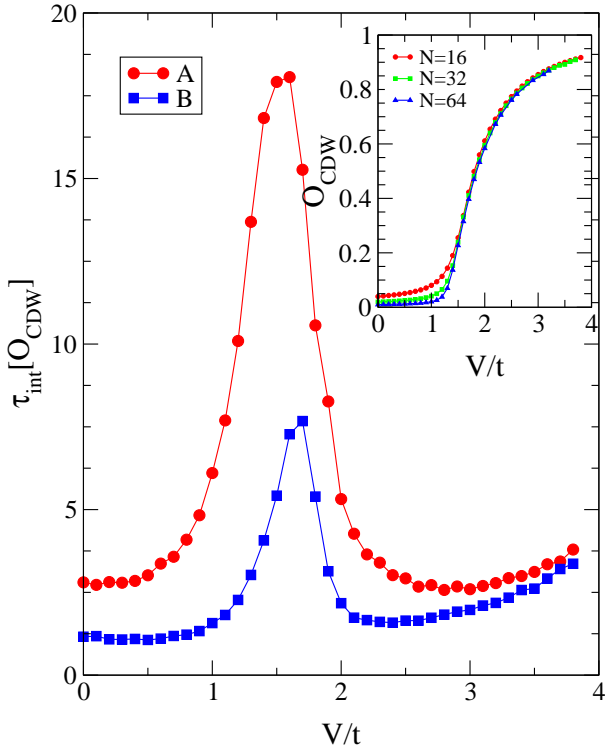}
\caption{Integrated autocorrelation times for the CDW order-parameter for the two solution A and B in a $N=16$ sites system at $\beta t =32$, $U/t=2$, $\mu=J_\perp=J_z=H_z=0$ as functions of $V$. The inset shows  ${\cal O}_{\rm CDW}(\pi)$ measured for system sizes $N=16$, $32$, $64$ at low temperatures $\beta t=2N$.
\label{fermicdw}}
\end{figure}   

Lets now consider another update type where two particles are added. 
Even though the update types considered above are sufficient for 
ergodicity, this update type is necessary if one wishes to measure 
superconducting correlation functions (local pairs). Then starting
from a diagonal vertex it is clear that one cannot reach an off-diagonal vertex. Thus there are only two possibilities, continuing straight through or bouncing.
To avoid bouncing the weights of the resulting vertices must be equal, which means $V=\mu=0$. Otherwise bounces are necessary.
Entering an off-diagonal vertex also leads to only two exit possibilities, but there both have weight $t$ and so the bounce probability can be set to zero.

\section{Spin-1/2 ferromagnet in a transverse field}
As an example of a model without a conservation law we consider the spin-1/2 $XXZ$ ferromagnet in a transverse magnetic field, that is a field along the $x$-direction. 
\bea
{\cal H} & = & -\sum_{<ij>} \left\{ \left( S^x_i S^x_j + S^y_i S^y_j
\right)                           - J_z S^z_i S^z_j +C\right\} \nonumber \\
	 &   & + h_x \sum_i S^x_i   
\eea
The exchange coupling in the spin XY-plane is restricted to be ferromagnetic as a $\pi$-rotation on one sublattice which was necessary to obtain an antiferromagnet in the zero field case cannot be employed here without introducing a minus sign coming from the $h_x$-term. Alternatively one can view this as an antiferromagnet in a {\em staggered} magnetic field.

\begin{figure}
\includegraphics[clip,width=7cm]{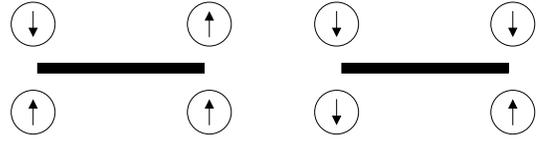}
\caption{Examples of vertices which do not conserve the total spin in the Z-direction. In the transverse field $s=1/2$ XXZ model all such vertices have weight $\bond{h}_x/2$. ($\bond{h}_x = h_x/Z$).
\label{transweight}}
\end{figure}

The transverse field introduces vertices where the sum of the spins on the lower two legs is not equal to the sum on the upper two legs, see Fig.~\ref{transweight}. 
This reflects the fact that $\sum_i S^Z_i$ is not a good quantum number in the presence of a transverse field. Thus the conservation law utilized in Sec.~IV cannot be used and we must include the possibilities of a state change on just one leg, the entrance or the exit leg, keeping the state on the other legs unchanged. The path construction in the absence of a conservation law always start by keeping the entrance leg on the first vertex unchanged, while it stops when the state on the exit leg is not changed. 

\begin{figure}
\includegraphics[clip,width=8cm]{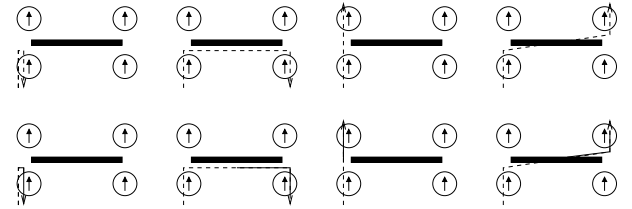}
\caption{The possible exit-legs and exit state changes when the entrance leg is the lower left leg on the vertex with all spins up and the entrance leg is unchanged. The dashed line indicates that the state should remain unchanged, while the solid line indicates a spin flip. 
\label{vertex8}}
\end{figure}

To find the directed loop equations here we first look at the vertex where all spins are up and the entrance leg is unchanged. Then we have the possibilities of exiting at one of the four legs as well as changing or not changing the state of the exit leg. In all there are eight possibilities, see Fig.~\ref{vertex8}, thus the directed loop equations have dimension 8. While the dimension is rather big there are only two different vertex weights, $W(_{\up\up}^{\up\up})$ and $W( _{\down \up}^{\up \up})$ (the vertices with one leg different from the others are all degenerate). 
Next one should take the same vertex, but now change the state on the entrance leg. Now there are just seven possibilities as changing both the lower legs leads to a zero-weight vertex. This procedure should be repeated for all vertices, entrance-legs and update types. 

It is interesting to note that the region where one can avoid bounces is in fact bigger in this representation using a transverse field than found in Sec.~IV.
Take the situation where the lower left leg on a diagonal vertex is changed. The vertices belonging to the equation set generated in this way are the vertices that would be generated without the transverse field {\em plus} the four vertices with just the in-leg changed and no change of state on the out-leg. The no-bounce-criterion is then (assuming that C is chosen such that the diagonal vertices are biggest)
\be
   \f{|J_z|}{2}  \leq  \f{1}{2} + 4\f{\bond{h}_x}{2}. 
\ee
One must also consider the sets of the type shown in Fig.~\ref{vertex8} where the in-leg is not changed. As there in these cases are just two different weights out of a total of eight the inequality (\ref{bfallowed}) is always satisfied.

As explained in Sec.~IV there is an ambiguity in solution B when many vertex weights are equal. Here we have used the convention that when two weights are equal the one with the lowest exit leg comes first. In the rare case where the exit legs are also equal, the one with the no state change on the exit leg comes first. We have tried solution A, B1 and a solution B2 which is particular for systems of dimensionality less than or equal to 8:
\bea
a_{12} & = & (W_1+W_2-W_3-W_4)/2 \nonumber \\
a_{13} & = & (W_1-W_2+W_3-W_4)/2 \nonumber \\
a_{23} & = & (-W_1+W_2+W_3+W_4)/2 \nonumber \\
a_{14} & = & W_4 - (W_5+W_6+W_7+W_8)/4 \nonumber \\
a_{15} & = & a_{45} = W_5/4  \nonumber \\
& \vdots & \\
a_{18} & = & a_{48} = W_{n}/4  \nonumber \\
a_{56} & = & (W_5+W_6-W_7-W_8)/4 \nonumber \\
a_{57} & = & (W_5-W_6+W_7-W_8)/4 \nonumber \\
a_{58} & = & W_8/2 \nonumber \\
a_{67} & = & (-W_5+W_6+W_7+W_8)/4 \nonumber \\
\eea
This solution reduces also to solution B when $W_5=W_6=\ldots = W_n = 0$ and is restricted to the region where $-W_5+W_6+W_7+W_8 \geq 0$.

We found that solutions A and B2 perform better than solution B1 in all cases studied. This can perhaps be explained by the fact that in B1 many processes vanish when the smaller weights are equal which is the case for small to intermediate fields. The difference in efficiency between solution A and B2 is small in the cases studied here. We have been unable so far to find a bounce-free solution which clearly outperforms solution A.

For $h_x=0$ the 2d XY-model ($J_z=0$) exhibits a phase-transition of the Kosterlitz-Thouless type\cite{HK2} as a function of temperature where the helicity modulus as measured by the second derivative of the free energy to a twist $\theta$ in the boundary conditions shows an emerging discontinuity with increasing system size at $T_c$. In zero field this quantity is efficiently measured as the fluctuations in the spatial winding number of the loops. With a magnetic field term one can still measure the second derivative of the free energy with respect to a twist in the boundary condition as fluctuations in a ``winding'' number provided one redefines the winding number to include a term coming from the magnetic field term in the Hamiltonian:
\be
  \f{d^2 F}{d\theta^2} = \f{T}{2} \left( \langle W_{h_x}^2 \rangle 
                                 + \langle W_{h_y}^2 \rangle \right) 
\ee
where we have symmetrized the expression in $x$ and $y$. The modified ``winding'' number $W_{h_\sigma}$, where $\sigma = x,y$, is
\bea
W_{h_\sigma} & = & \f{1}{N_\sigma} \sum_{p} \left\{ 
 \left( \delta_{p, _{\up \down}^{\down \up}}
-\delta_{p, _{\down \up}^{\up \down}} \right) \delta_{\vec{p},\hat{\sigma}}
 \right.  \\
&   &
+ \sigma(p) \left[ 
  \delta_{p, _{\down \up  }^{\up   \up}}
 -\delta_{p, _{\up   \up  }^{\down \up}}
 +\delta_{p, _{\down \down}^{\up \down}}
 -\delta_{p, _{\up \down  }^{\down \down}} \right.
    \nonumber \\
&   &
  \left. \left. 
 +\delta_{p, _{\up \down  }^{\up \up  }}
 -\delta_{p, _{\up   \up  }^{\up \down }}
 +\delta_{p, _{\down \down}^{\down \up  }}
 -\delta_{p, _{\down \up  }^{\down \down}} \right]  \right\} \nonumber
\eea
where the sum is over all vertices $p$ in the linked list. The
Kronecker $\delta$-functions contribute whenever the vertex $p$ equals
the indicated vertex. $\sigma(p)$ is the $\sigma$-coordinate of the site whose state is changed in the vertex $p$. The symbol $\delta_{\vec{p},\hat{\sigma}}$  means that the bond to which the vertex is attached must lie in the $\sigma$-direction, and $N_\sigma$ is the number of sites in the $\sigma$-direction.     
\begin{figure}
\includegraphics[clip,width=8cm]{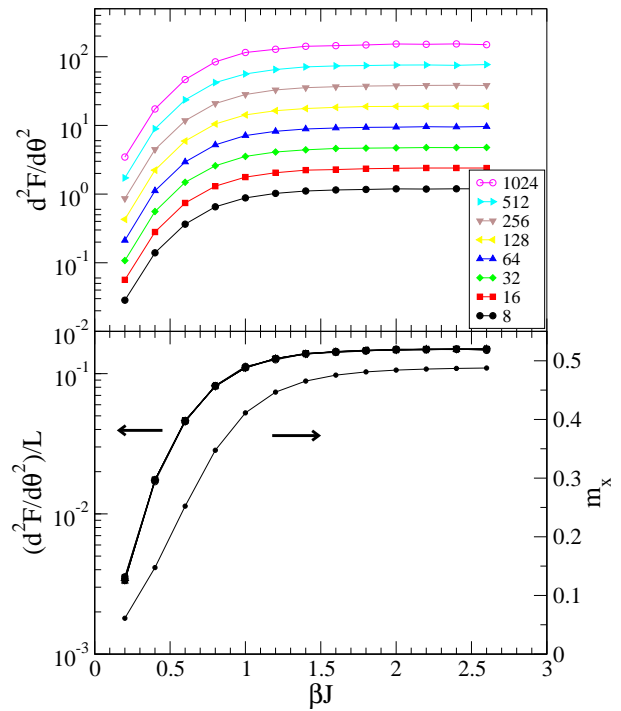}
\caption{The second derivative of the change in free energy resulting from a twist in the boundary conditions are shown (top panel) as functions of inverse temperature for different square lattice system sizes $L \times L$ ranging from $L=8$ to $L=1024$. The bottom panel shows the plot of $\partial^2 F/\partial \theta^2 L$ which collapses onto one curve for all $L$ as well as the magnetization obtained using the biggest system size (L=1024).}
\label{spinstiff}
\end{figure}    

In Fig.~\ref{spinstiff} we have measured $\partial^2 F/\partial \theta^2$ at $h_x=1$ as a function of (inverse) temperature for $L \times L$ square lattices where $L$ ranges from $8$ to $1024$.  
For strong fields the spins are predominantly in the $x$-direction, and it is expected that a twist in the boundary condition will cause $\partial^2 F/\partial \theta^2$ to depend linearly on $L$. In the bottom panel we confirm this by showing how the scaled curves(scaling factor $1/L$) collapse onto one curve for all $L$'s. In this plot we also show the magnetization for the biggest system size.

\section{Discussion}
We have shown how to construct probability tables for the directed loop
update in the SSE method in a model-independent way. 
The construction involves solutions of the directed loop equations. These equations arise from considering vertices that can be reached from a certain vertex when states on two legs of that vertex are changed. For a given model there are many such equation sets. To cope with the many sets it is best to generate them in the set-up part of the Monte Carlo program. Efficiency is not an issue in this construction process as the {\em same} probability tables will be employed throughout the simulation. To construct the probability tables the program should
go through every entrance leg on every vertex type occurring in the model under consideration. For each of these entrance legs the program also have to consider every possible update, as there should be one probability table associated with every entrance leg, vertex and update type. For each of these entrance legs and update types one finds all possible exit legs and exit states, and thus the related vertices. The number of vertices with non-zero weights reached in this way determines the effective dimensionality of the directed loop equations. Having found this dimension and weights for the different vertices one can then pick the probabilities of
moving from vertex $i$ to $j$ as $P(i \to j) = a_{ij}/W_i$, where $a_{ij}$ is
gotten from the general solutions described in Sec.~III.

The solutions are naturally divided into two classes; those with and those without bounces: processes where the loop back-tracks along its path. Bounces should generally be avoided as they are inefficient. The precise criterion for when bounces can be avoided is given in Eq.~(\ref{bfallowed}). When this
criterion is not fulfilled, bounces are necessary. A general solution which then always can be applied is given in Eq.~(\ref{genbounce}) where there is only one bounce, namely bouncing off the vertex with the biggest weight.  

Whenever bounces can be avoided it is likely that there exists a bounce-free algorithm which is more effective. This is supported by the results in\cite{SS} and in the examples considered here with the exception of the transverse field XXZ model where we have not been able to find a bounce-free solution which clearly is more efficient than solution A. However the author believes that such a solution exists. 

For directed loop equation sets with dimension $\leq 3$, as is the case for all the equation sets in the $s=1/2$ XXZ model as well as for the 1D spinful fermion model considered here, the bounce-free solution is unique and is given by Eq.~(\ref{bouncefree3}). For sets with dimension $> 3$ the bounce-free solution is not unique. In Eq.~(\ref{parbouncefree}) we have parametrized all bounce-free solutions in the 4-dimensional case. Testing the efficiency of different parameter choices on a model where there is only one 4 dimensional equation set, the s=1 Heisenberg model in zero field, we find that even in the case where all four weights related by the set are equal, the most effective solution is not the most symmetric one (all off-diagonal $a_{ij}$'s of equal magnitude). Thus we conclude that it is not possible to find the most effective solution based on the weights of a single isolated directed loop equation set alone. This is natural as the efficiency depends on the overall loop motion through all possible vertices and not just the motion within a certain subset of vertices related by the same directed loop equation set. Our finding of the most efficient algorithm for the spin-1 model coincides with the most efficient direct algorithm found by Harada and Kawashima\cite{HK}.

We have also shown how the directed loop equations can be applied to study spin-s XXZ models, lattice bosons, 1D spinful fermions and transverse field spin models using the same computer code just changing the number of allowed states on each site as well as the vertex weights. We have also worked out expressions for the regions where one can construct algorithms without bounces for these models.
       
While finding the most efficient solution to the directed loop equations among the many possible ones remains a difficult task, the solutions given here are generally more efficient than the solution employed in Ref.\cite{loopsandvik}. 
The many possible solutions to the directed loop equations should be seen as an asset. They are powerful tools allowing efficient simulations of a wide class of quantum models.

\begin{acknowledgments}
The author would like to thank Anders Sandvik for valuable discussions.
Monte Carlo calculations were in part carried out using NorduGrid,
a Nordic Testbed for Wide Area Computing and Data Handling. 
\end{acknowledgments}

\appendix

\section{The four-dimensional no-bounce solution when two or more weights are equal \label{Appendix}}
The no-bounce solutions in the four dimensional case can be parametrized as
\bea
	a_{12} & = & (W_1+W_2-W_3-W_4)/2+a_{34} \nonumber \\
	a_{13} & = & (W_1-W_2+W_3-W_4)/2+a_{24} \label{parbouncefree2} \\
	a_{23} & = & (-W_1+W_2+W_3+W_4)/2-\left( a_{34}+a_{24} \right) \nonumber \\
	a_{14} & = & W_4 -\left( a_{34}+a_{24} \right) \nonumber \\
\eea
where it is assumed that $W_1 \geq W_2 \geq W_3 \geq W_4$. When
two or more of these weights are equal the ordering is ambiguous.
However, any solution obtained for a chosen ordering and choices 
of $a_{24}=a$ and $a_{34}=b$ is  identical to a solution obtained
using another ordering and other values of $a_{24}$ and $a_{34}$. The
purpose of Table~\ref{table} is to relate values of $a_{24}$ and $a_{34}$ for
different orderings. 
We choose as a reference ordering the order ABCD which means that $W_1=W_A$,
$W_2=W_B$, $W_3=W_C$ and $W_4=W_D$. As an example consider the case where
$W_C=W_D$ but $W_A > W_B > W_C$. Then the two orderings ABCD and ABDC are
inequivalent. From the third entry in Table~\ref{table} we read that a choice
of $a_{24}=a$ and $a_{34}=b$ for the ordering ABCD give the same rules
as the ordering ABDC with the rules $a_{24}=W_\Delta-a-b$ and $a_{34}=b$.
$W_\Delta = (-W_1+W_2+W_3+W_4)/2$.
\begin{table}
\caption{\label{table}
The table shows how the solution of Eqs.(\ref{parbouncefree2}) with ordering ABCD and $a_{24}=a$, $a_{34}=b$
are related to other solutions with different orderings of vertices when two or more weights are equal. $W_\Delta = (-W_1+W_2+W_3+W_4)/2$. 
($^*$ when all weights are equal the transformations ($1 \leftrightarrow 2$,$3 \leftrightarrow 4$), ($1 \leftrightarrow 3$,$2 \leftrightarrow 4$) and
($1 \leftrightarrow 4$,$2 \leftrightarrow 3$) are symmetry transformations implying in particular that the orderings ABCD,BADC,CDAB and DCBA give the same rules for a particular choice of $a$ and $b$.) 
}
\begin{ruledtabular}
\begin{tabular}{llll}
                   & Order  & $a_{24}$       & $a_{34}$ \\
$W_A=W_B>W_C>W_D$  & $\rm{BACD}$   & $W_D-a-b$      & $b$  \\
$W_A>W_B=W_C>W_D$  & $\rm{ACBD}$   & $b$            & $a$  \\
$W_A>W_B>W_C=W_D$  & $\rm{ABDC}$   & $W_\Delta-a-b$ & $b$  \\
$W_A=W_B>W_C=W_D$  & $\rm{BACD}$   & $W_D-a-b$      & $b$  \\
                   & $\rm{ABDC}$   & $W_D-a-b$      & $b$  \\
                   & $\rm{BADC}$   & $a$            & $b$  \\
$W_A=W_B=W_C>W_D$  & $\rm{BACD}$   & $W_D-a-b$      & $b$  \\
                   & $\rm{BCAD}$   & $b$            & $W_D-a-b$ \\
                   & $\rm{ACBD}$   & $b$            & $a$  \\
                   & $\rm{CABD}$   & $W_D-a-b$      & $a$  \\
                   & $\rm{CBAD}$   & $a$            & $W_D-a-b$ \\
$W_A>W_B=W_C=W_D$  & $\rm{ABDC}$   & $W_\Delta-a-b$ & $b$  \\
                   & $\rm{ACBD}$   & $b$            & $a$  \\
                   & $\rm{ACDB}$   & $W_\Delta-a-b$ & $a$  \\
                   & $\rm{ADBC}$   & $b$            & $W_D-a-b$ \\
                   & $\rm{ADCB}$   & $a$            & $W_D-a-b$ \\     
$W_A=W_B=W_C=W_D$  & $\rm{ABCD}^*$ & $a$            & $b$  \\
                   & $\rm{BACD}^*$ & $W_\Delta-a-b$ & $b$  \\
                   & $\rm{CBAD}^*$ & $a$            & $W_\Delta-a-b$  \\ 
                   & $\rm{DBCA}^*$ & $b$            & $a$  \\
                   & $\rm{ACDB}^*$ & $W_\Delta-a-b$ & $a$  \\
                   & $\rm{ADBC}^*$ & $b$            & $W_\Delta-a-b$ 
\end{tabular}
\end{ruledtabular}
\end{table}

\end{document}